\documentclass[11pt]{article}

\usepackage{a4wide}
\usepackage{amsmath,amssymb,amsthm,amscd}
\usepackage[pdftex]{graphicx}

\newcommand{\bt}{\pmb\theta}
\newcommand{\vc}[1]{{\bf #1 }}

\def\bothidenty{\rlap{\hbox to.97\wd0{\hss\vrule height.06\ht0 width.82\wd0}}
 \copy0\rlap{\kern-.36\wd0\vrule height1.05\ht0 width.05\ht0}\kern.14\wd0}

\begin{document}

\title{Boosting Bayesian Parameter Inference of Nonlinear Stochastic Differential Equation Models by Hamiltonian Scale Separation}
\author{Carlo Albert\footnote{Eawag, Swiss Federal Institute of Aquatic Science and Technology, 8600 D\"ubendorf, Switzerland.}, Simone Ulzega\footnotemark[1] and Ruedi Stoop\footnote{Institute of Neuroinformatics and Institute of Computational Science UZH/ETHZ, Irchel Campus 8057 Zurich, Switzerland.}}

\maketitle

\abstract
Parameter inference is a fundamental problem in data-driven modeling.
Given observed data that is believed to be a realization of some parameterized model, the aim is to find parameter values that are able to explain the observed data.
In many situations, the dominant sources of uncertainty must be included into the model, for making reliable predictions. This naturally leads to stochastic models.
Stochastic models render parameter inference much harder, as the aim then is to find a distribution of likely parameter values. In Bayesian statistics, which is a consistent framework for data-driven learning, this so-called posterior distribution can be used to make probabilistic predictions.
We propose a novel, exact and very efficient approach for generating posterior parameter distributions, for stochastic differential equation models calibrated to measured time-series.
The algorithm is inspired by re-interpreting the posterior distribution as a statistical mechanics partition function of an object akin to a polymer, where the measurements are mapped on heavier beads compared to those of the simulated data.
To arrive at distribution samples, we employ a Hamiltonian Monte Carlo approach combined with a multiple time-scale integration.
A separation of time scales naturally arises if either the number of measurement points or the number of simulation points becomes large.
Furthermore, at least for 1D problems, we can decouple the harmonic modes between measurement points and solve the fastest part of their dynamics analytically.
Our approach is applicable to a wide range of inference problems and is highly parallelizable.\\

\maketitle

\section{Introduction}

\begin{figure}
   \centerline{\includegraphics[width=\textwidth]{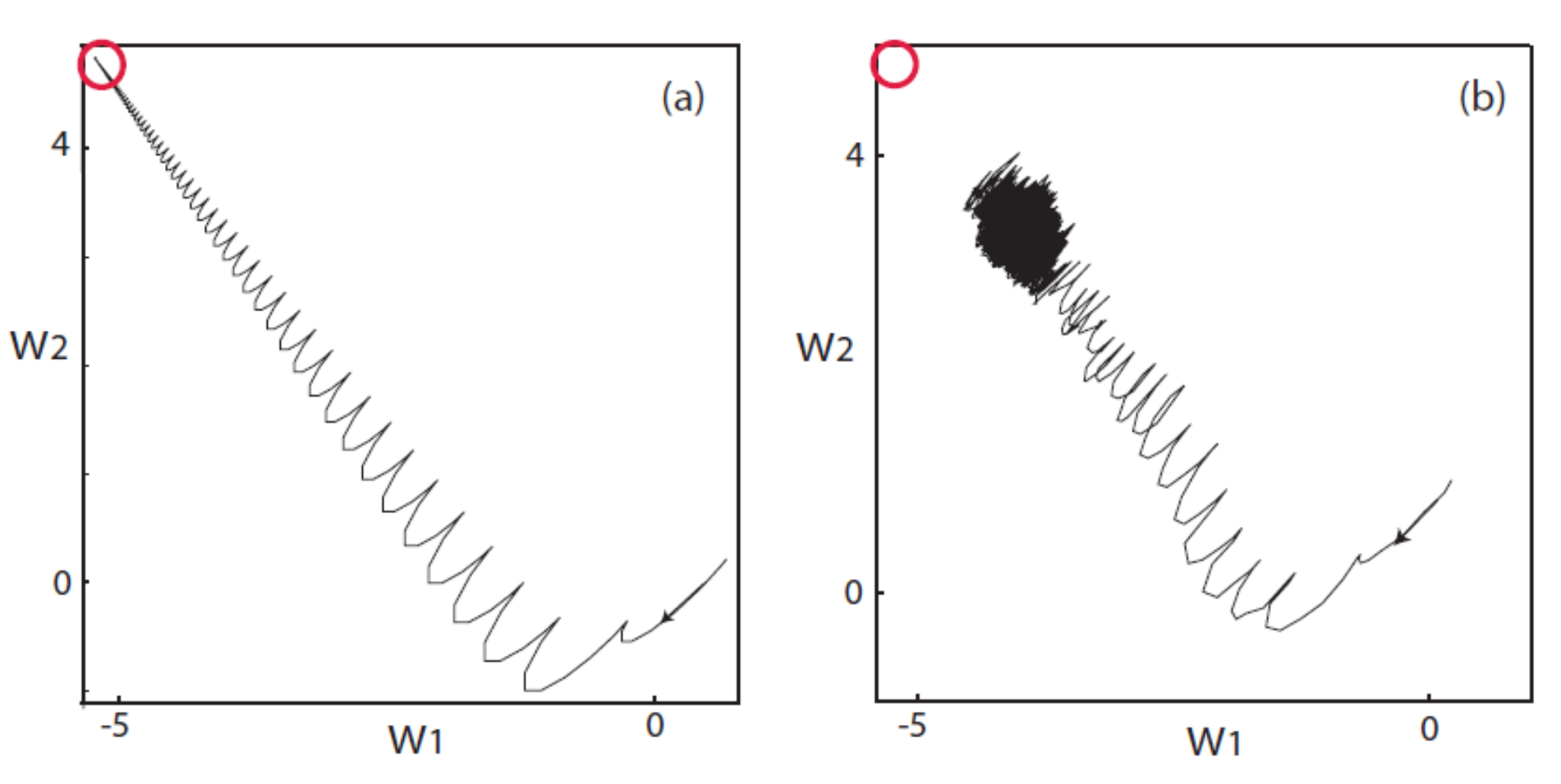}}
    \caption{General problem setting: Parameter estimation (synaptic weights $w_1$,$w_2$) of a perceptron, from a) noiseless, b) noisy data (noise sampled from a flat distribution over the  interval $[-0.2,0.2]$). Whereas for noiseless data the estimates perfectly converge, for noisy data the estimates  the system attempts to also include the noise, leading to a nontrivial distribution of the parameter estimates, and rendering the extraction of the optimal parameters a nontrivial task. Open circles: location of the optimal parameters in the noiseless case.}
    \label{neural}
\end{figure}

Modeling a dynamical process starts with a basic model that is usually obtained from a more or less deep insight into the nature of the process. The next step is the determination of the parameters of the model, based on observed data, which is generally a highly nontrivial task, in particular when complex behavior of such systems needs to be predicted, or when the measurements are noisy.
A minimal example is a perceptron \cite{rosenblatt_1958_perceptron, steeb_2004_problems}, the basic element of a neuronal network, that predicts the double cosine value associated with the input of the corresponding sine function plus the sine's value at a fixed earlier time. While this task can easily be achieved for clean data using gradient descent learning, for noisy input data, this is largely impossible, as noise cannot be learned. The result is a distribution of potential parameter values (Fig.~(\ref{neural})).

In Bayesian statistics, knowledge about parameters is expressed by probability distributions and learning is implemented as an update rule on these distributions (see, e.g. \cite{box_2011_bayesian}).
If a constant noise term is added to the output of a deterministic model, such as in the perceptron example above, Bayesian inference is straightforward.
If noise enters the formulation of the model equations, however, Bayesian inference all of a sudden becomes computationally very expensive.

In our paper, we demonstrate the calibration of ordinary 1D stochastic differential equation (SDE) models based on noisy time series, and the quantification of the resulting parametric uncertainty. The generic approach that we use is exemplified by a simple SDE model from hydrology.

Problems of this kind are commonly solved by Monte Carlo (MC) methods that are based on simulating model realizations and comparing them to the data. Popular methods are particle filters \cite{chopin_2013_SMC2, liu_2012_filters}, Metropolis-within-Gibbs algorithms \cite{tomassini_2009_smoothing, reichert_2009_timedepParameters} or Approximate Bayes Computations \cite{marin_2012_ABC, Albert_2014_ABC, toni_2009_ABC, toni_2010_ABC}.
A major problem with these simulation-based methods is, however, their inefficiency in the presence of many data points or high dimensions. One solution is to map the output space to a smaller dimensional space of summary statistics, and accept/reject proposed model parameters depending on how well associated model runs conform with the data in terms of these summary statistics \cite{Fearnhead_2012_ABC}. However, how to choose the summary statistics to achieve a significant representation of the posterior parameter distribution is a largely unsolved problem.

These difficulties can be remedied with a reinterpretation of the Bayesian posterior distribution as the partition function of a statistical mechanics system and by simulating the dynamics of the latter.
After discretizing the time of the original problem, we are led to a problem akin to the statistical mechanics of a polymer 
with harmonic bonds in an exterior potential \cite{chandler_1981_polymer}. In this framework, the measurements are interpreted as an additional exterior potential that acts only on the polymer's `measurement beads' and confines their dynamics within the measurement uncertainty.
The model parameters are interpreted as additional degrees of freedom coupling to all the beads of the polymer.
To simulate the dynamics of this system, we apply the Hamiltonian Monte Carlo (HMC) algorithm \cite{duane_1987}, which combines Molecular Dynamics \cite{alder_1959_MD, rahman_1964_MD} with the Metropolis algorithm \cite{metropolis_1953}.
Compared to traditional methods, the use of the Hamiltonian approach achieves much higher acceptance rates since data points are already used for the suggestion of new parameters, and thus model realizations incompatible with the data are never considered. The drawback is that the model equations need to be known and derivatives have to be calculated.

HMC requires two sets of parameters to be tuned: (i) the parameters that define the kinetic energy of the statistical mechanics system and (ii) the parameters that define the numerical integration scheme of Hamilton's equation in the molecular dynamics part of the HMC algorithm. Efficiency of HMC algorithms can be gained if the kinetic term is made dependent on the configuration geometry of the statistical mechanics system. If the Riemann geometry of the parameter space of statistical models is taken into account, then the simulated search of paths across this manifold samples the target density in an utmost efficient way \cite{girolami_2011_HMC}. Unfortunately, this procedure is both demanding and computationally costly,  depending strongly on the quality of the space's extracted geometry.


Here we
explore a computationally simpler approach, which, to our knowledge, has never been applied in the context of Bayesian inference before.
Depending on the number of discretization points needed to approximate the original SDE system and the number of measurement points, the dynamics of the statistical mechanics system happen on very different time scales. This suggests a multiple time scale integration technique for the simulation of the statistical mechanics system \cite{tuckerman_1993}.
We will show that for 1D SDE we can always find a parametrization, which decouples the harmonic modes in between measurement points from both the measurement points and the model parameters and allows for an analytical time-saving solution of the fastest part of the dynamics.
Whilst for higher dimensional problems it is not always possible to solve part of the dynamics analytically, we believe that scale separation alone will render many SDE amenable to a full-fledged Bayesian inference with time-series.
In fact, scale separation appears to be a generic feature if the dynamics of the SDE requires a large number of discretization points.

\section{Inference Problem Setting}

Consider, for simplicity and concreteness, a reservoir dynamics $S(t)$ that on the observation time-scale is linear, with other (inflow and outflow) processes happening at much shorter time scales, so that they can be described by white noise. Furthermore, assume that this noise scales linearly with the system state $S(t)$. The model equation is thus given by the SDE

\begin{equation}\label{sde}
\dot{S}(t) = r(t) - \frac{1}{K}\left(1+\frac{\gamma}{2}\right) S(t)
+
\sqrt{\frac{\gamma}{K}} S(t){\eta}(t)\,,
\end{equation}
where $r(t)$ denotes the time varying rain input, $K$ denotes the retention time, $\gamma$ is the noise strength, and $\eta(t)$ indicates the white noise property, i.e.,
\begin{equation}\label{whitenoise}
\langle\eta(t)\eta(t')\rangle = \delta(t-t')\,.
\end{equation}
Eq.~(\ref{sde}) is to be understood in the Stratonovich sense \cite{stratonovich_1968}.

Properties of a transformed version of (\ref{sde}) have been derived, for constant input \cite{dutre_1977_SDE, schenzle_1979_multStochProc, fujisaka_1986_intermittency}.
Here, suffice it to say that, for constant input $r(t) = r_0$, the equilibrium distribution, $P_{eq}(S)$, is an inverse gamma distribution with scale parameter $2Kr_{0}/\gamma$ and shape parameter $(2+\gamma)/\gamma$ (see Sect. \ref{Sect:inference}), i.e.,
\begin{equation}\label{inverse_gamma}
  P_{eq}(S)
  \propto
  S^{-2(1+\gamma)/\gamma}e^{-2Kr_{0}/(\gamma S)}\,.
\end{equation}
The mean of this expression equals the equilibrium solution of the unperturbed system ($\gamma=0$)
$\langle S\rangle_{eq}=Kr_{0}$
and its variance, for $\gamma< 2$, is given by $ \langle (S - \langle S\rangle_{eq})^2\rangle_{eq}
  =
  K^2r_{0}^2
  \gamma/(2-\gamma)$, which, for $\gamma\geq 2$, is seen to diverge.
The power-law decay of the inverse gamma distribution is reminiscent of the invariance of Eq.~(\ref{sde}) under re-scaling of both $r(t)$ and $S(t)$. In real-world hydrology, for which Eq.~(\ref{sde}) is a model, indeed often fat-tailed error distributions are observed \cite{thyer_2009_fattails}.

While this equation was motivated by a popular hydrological model \cite{breinholt_2011_SDE, livina_2003_dischargeFluctuations}, it is by no means restricted to this context. By means of the transformation $S(t) = 1/n(t)$, Eq.~(\ref{sde}) turns into a model that has been suggested, e.g., as a phenomenological description of the dynamics of the neutron density in nuclear reactors \cite{dutre_1977_SDE}.

In our setting, the input $r(t)$ is a smooth and nowhere vanishing function. We assume the observed time-series, $y_s$, to be the outflow of the reservoir, $S(t)/K$, observed at times $0=t_1<t_2<\dots < t_{n+1}=T$, with multiplicative independent log-normal errors,

\begin{equation}\label{data}
  \ln \left( y_s \right)
  =
  \ln \left( \frac{S(t_s)}{K} \right)
  +
  \sigma\epsilon_s\,,\quad s=1,\dots,n+1\,,
\end{equation}
where the $\epsilon_s$ are uncorrelated standard normal errors.
For simplicity, we assume $\sigma$ as well as the input $r(t)$ to be known, so that we are left with the task of inferring parameter combinations $(K,\gamma)$ that are compatible with the data given by Eq.~(\ref{data}) in the Bayesian sense, where knowledge about parameters is expressed in terms of probability distributions.

We start our treatise by assuming that we have prior knowledge about a parameter vector, $\bt$, in the form of a probability distribution, $f_{\mbox{\small prior}}(\bt)$, and measured data, $\vc y$, believed to be a realization of the model.
The posterior knowledge, combining prior knowledge with the one acquired from data, is calculated by means of equation
\begin{equation}\label{Bayes}
  f_{\mbox{\small post}}(\bt|\vc y)
  =
  \frac{f_{\mbox{\small prior}}(\bt)L(\vc y|\bt)}
  {\int f_{\mbox{\small prior}}(\bt')L(\vc y|\bt')d\bt'}\,,
\end{equation}
where $L(\vc y|\bt)$ is the probability distribution for model outputs given model parameters, evaluated at the measured data (the infamous {\em likelihood function}).

Before we set out to derive from Eqs. (\ref{sde}), (\ref{whitenoise}) and (\ref{data}) the likelihood function, we express the parameters and state variables by dimensionless quantities.
Due to scale-invariance of the noise term, $\gamma$ is already dimensionless. State variable $S(t)$ and parameter $K$ are replaced by the dimensionless quantities $q(t)$ and $\beta$, respectively, which are defined by the transformations $\beta=\sqrt{T\gamma/K}$ and $S(t)=(T\gamma r(t)/\beta^2)e^{\beta q(t)}$.
In these new variables and parameters, Eq.~(\ref{sde}) becomes the nonlinear SDE with constant noise
\begin{equation}\label{standardform}
  \dot q(t)
  =
  \frac{\beta}{T\gamma}e^{-\beta q(t)}
  -
  \frac{1}{T}\rho(t)
  +
  \frac{1}{\sqrt{T}}\eta(t)\,,
\end{equation}
with $\rho(t)  =  (T/\beta)(\mbox d/\mbox{dt})\left[ \ln \left( r(t) \right) \right]
  +
  (2+\gamma)\beta/(2\gamma)$.

The probability $P(q_1,T|q_0,0)$ of finding the system in a state $q_1$ at time $t = T$ if it was in an initial state $q_0$ at time $t = 0$, is expressed as a {\em path-integral} as
\begin{equation}\label{pathint}
P(q_1,T|q_0,0)
=
\frac{1}{Z}
\int
e^{-{\mathcal S}[q,\dot q]}
\delta(q(T)-q_1)
\delta(q(0)-q_0)
\mathcal{D}q \,,
\end{equation}
where the integral extends over all paths $q:[0,T]\rightarrow \mathbb R$ and where the path-measure $\mathcal Dq$ is formally written as the infinite product
${\mathcal Dq}=\prod_{t}dq(t)$.
The {\em action} is a functional on the space of paths and reads \cite{lau_2007}
\begin{equation}\label{action1}
{\mathcal S}[{q},\dot q]
=
\frac{1}{T}
\int_0^T dt \left\{
\frac{1}{2}
\left(
    T\dot q(t)
    +
    \rho(t)
    -
    \frac{\beta}{\gamma}e^{-\beta q(t)}\right)^2
    -
    \frac{\beta^2}{2\gamma}e^{-\beta q(t)}
\right\} \,.
\end{equation}
This action includes the Jacobian that is introduced when changing coordinates from
 ${\eta(t)}$ to $q(t)$.

If we denote the parameter vector $\bt=(\beta,\gamma)^T$ and assume a flat prior, the posterior (\ref{Bayes}) is, as a function of $\bt$, proportional to the likelihood function
\begin{equation}\label{posterior_pathint}
  f_{\mbox{\small post}}(\bt | \vc y)
  \propto
  \int
  \exp\bigg[
    -\frac{1}{2}
    \sum_{s=1}^{n+1}
    \frac
    {(\ln(y_s/r(t_s))-{\beta q(t_s)})^2}
    {\sigma^2} -{{\mathcal S}}[{q},\dot q]
  \bigg]
  {\mathcal Dq}
  \,.
\end{equation}
Whereas the first term in the exponent describes the log-probability distribution of model outputs, for given model parameters, inputs and a system realization $q(t_s)$, the second term is the log-probability of the associated system realization $q(t)$.

When applying this approach now to real-world problems, instead of undertaking a prohibitive numerical computation of the path integral, we apply HMC to sample parameter vectors from a joint distribution of system realizations and model parameters given by an appropriate discretization of the action of the path-integral. By doing so, we observe that we obtain distinct regimes of time scales in the Hamiltonian that can be separated (see Sect. \ref{Sect:timescale}). This time scale separation simplifies and boosts our algorithm; in many cases it even permits parts of the required integrations to be done analytically.

\section{Algorithm}

\subsection{Inference algorithm}\label{Sect:inference}

For the inference algorithm, it is necessary to rewrite action~(\ref{action1}) with the help of the time-dependent potential $U(q,t)= \frac{1}{\gamma}e^{-\beta q}+q\rho(t)$ as
\begin{multline}\label{action}
{\mathcal S}[{q},\dot q]
= \frac{1}{T}
\int_0^T dt\bigg\{
    \frac{1}{2}
    T^2\dot q^2(t) +
    \frac{1}{2}
    \left(\rho(t)-\frac{\beta}{\gamma}e^{-\beta q(t)}\right)^2
    -
    T\frac{\partial U(q,t)}{\partial t}
    -
    \frac{\beta^2}{2\gamma}e^{-\beta q(t)}
\bigg\}
\\
+ U(q(T),T) - U(q(0),0)
\\
= \frac{1}{T}
\int_0^T dt\bigg\{
    \frac{1}{2}
    T^2\dot q^2(t) +
    \frac{1}{2}
    \left(\rho(t)-\frac{\beta}{\gamma}e^{-\beta q(t)}\right)^2
    -
    Tq(t)\dot\rho(t) -
     \frac{\beta^2}{2\gamma}e^{-\beta q(t)}
\bigg\}
\\
+
    \frac{1}{\gamma}e^{-\beta q(T)}+q(T)\rho(T)
   -\frac{1}{\gamma}e^{-\beta q(0)}-q(0)\rho(0)
\,.
\end{multline}

With the action in this form, we easily derive the equilibrium distribution, for constant input $r(t)=r_0$, by plugging (\ref{pathint}) and (\ref{action}) into the detailed balance condition
\begin{equation}\label{detailed_balance}
P(q_1 t_1 | q_0 t_0 ) P_{eq}(q_0) = P(q_0 t_1 | q_1 t_0 ) P_{eq}(q_1) \,,
\end{equation}
and using the transformation $q(t) \rightarrow q(-t)$.
We get, since $\dot\rho(t)= 0$,
$$
P_{eq}(q)
  \propto
  e^{-2U(q)}\,.
$$
Back transformation to the original variables leads to Eq. (\ref{inverse_gamma}).

For efficiently drawing parameter samples from (\ref{posterior_pathint}), we interpret the latter as the partition function of a 1D statistical mechanics system and simulate its dynamics employing the HMC algorithm \cite{duane_1987}. The model parameters  $\bt$ are interpreted as additional dynamical degrees of freedom coupling to the system variables $q(t)$. Each degree of freedom, $q(t)$ and $\bt$, is paired with a conjugate variable, $p(t)$ and ${\pmb\pi}$ respectively, so that the system is defined by the Hamiltonian

\begin{equation}\label{Hamiltonian}
    \mathcal{H}_{\mbox{\tiny HMC}}(q,\bt; p,{\pmb\pi})
    =
    K( p,{\pmb\pi}) + V( q,\bt)\,,
\end{equation}
where
\begin{equation}\label{K}
   K( p,{\pmb\pi})
   =
   \int_0^T \frac{ p^2(t)}{2m(t)}dt
   + \sum_{\alpha=1}^2\frac{\pi_\alpha^2}{2m_\alpha}\,,
\end{equation}
and $V( q,\bt)$ is the negative logarithm of the kernel of (\ref{posterior_pathint}).
The posterior (\ref{posterior_pathint}) can then
be expressed by the phase space path integral
\begin{equation}\label{phaseSpacePathInt}
    f_{\mbox{\small post}}(\bt | \vc y)
  \propto
  \int
  e^{-\mathcal {H}_{\mbox{\tiny HMC}}(q,\bt; p,{\pmb\pi})}
  {\mathcal Dp}
   {\mathcal Dq}
   d{\pmb\pi}
  \,.
\end{equation}

The HMC method, as a combination of the {\em Metropolis algorithm} \cite{metropolis_1953} and {\em molecular dynamics} methods \cite{alder_1959_MD, rahman_1964_MD}, iterates the following steps:
\begin{enumerate}
  \item
  Momenta $p(t)$ and ${\pmb\pi}$ are sampled from the Gaussian distributions defined by Eq.~(\ref{K}).
  \item
  The system is then allowed to evolve in $\left(q,\bt; p,{\pmb\pi}\right)$-phase space for an arbitrary time interval $\tau$ according to a volume-preserving and time-reversible solution of a discretized set of Hamilton equations.
  \item
  The discretization error on the energy preservation due to the previous step is corrected by a Metropolis acceptance/rejection step.
\end{enumerate}
The last step is the standard Metropolis algorithm, while the first two steps permit arbitrarily large jumps in phase space, while maintaining an arbitrarily large acceptance rate. Each new phase space configuration is associated with a combination of model parameters $\bt$, which is compatible with the data in the Bayesian sense. Thus, omitting a possible burn-in, the parameter marginal of the simulated Markov chain of configurations represents a sample of the posterior probability distribution.

In order to simulate the dynamics of the Hamiltonian (\ref{Hamiltonian}), we first need to discretize the path-integral (\ref{phaseSpacePathInt}).
Let us assume that the measurement time points $\left\{ y_s \right\}_{s=1,\dots, n+1}$ of the time series (\ref{data}) are equidistantly distributed on the time interval $[0,T]$, with $t_1=0$ and $t_{n+1}=T$.
Each interval between two consecutive data points is further partitioned into $j$ bins, such that we have a total of $nj+1=N>>1$ discretization points.
The path-integral (\ref{phaseSpacePathInt}) is then approximated by an ordinary integral, with the approximate path-measure $  \mathcal Dp\mathcal Dq  \approx  \prod_i dp_i dq_i$.
The discretized versions of $K( p,{\pmb\pi})$ and $V( q,\bt)$ are now given by
\begin{align}
   K( p,{\pmb\pi}) 
   &\approx
   \sum_{i=1}^N
   \frac{ p_i^2}{2m_i}\Delta t + \sum_{\alpha=1}^2\frac{\pi_\alpha^2}{2m_\alpha}\,,\label{Kdisc}
   \\
  V(q,\bt)  
  &\approx 
  \frac{\Delta t}{T} \sum_{i=2}^{N}
   \bigg\{ \frac{1}{2} T^2 \dot q_i^2 + \frac{1}{2}
     \left( \rho_i-\frac{\beta}{\gamma}e^{-\beta q_i} \right)^2
     -
    \frac{\beta^2}{2\gamma} e^{-\beta q_i} - T q_i\dot\rho_i \bigg\}\nonumber
    \\
  &+ 
  \frac{1}{\gamma} e^{-\beta q_N} + q_N \rho_{N} - \frac{1}{\gamma} e^{-\beta q_1} -  q_1 \rho_{2}
  +
  \sum_{s=1}^{n+1} \frac{(\ln(y_s/r_{(s-1)j+1}) - {\beta q_{(s-1)j+1}})^2}{2\sigma^2}\,,\label{Vdisc}
\end{align}
with $\dot q_i = (q_i-q_{i-1})/\Delta t$, $\rho_i = T \ln(r(t_{i})/r(t_{i-1}))/(\beta\Delta t)+(2+\gamma)\beta/(2\gamma)$ and $\dot\rho_i = (\rho_i-\rho_{i-1})/\Delta t$, and where terms of order $\mathcal O(N^{-1/2})$ were neglected.
Note that we did not apply the mid-point discretization that is associated with the Stratonovich convention. In Eq. (\ref{Kdisc}) this leads to a different dynamics that does not, however, alter the posterior we are interested in, and in Eq. (\ref{Vdisc}) we produce errors of the order $\mathcal O(N^{-1/2})$ that we neglect.

\begin{figure}[htb!]
    \centering
    \includegraphics[width=0.48\textwidth]{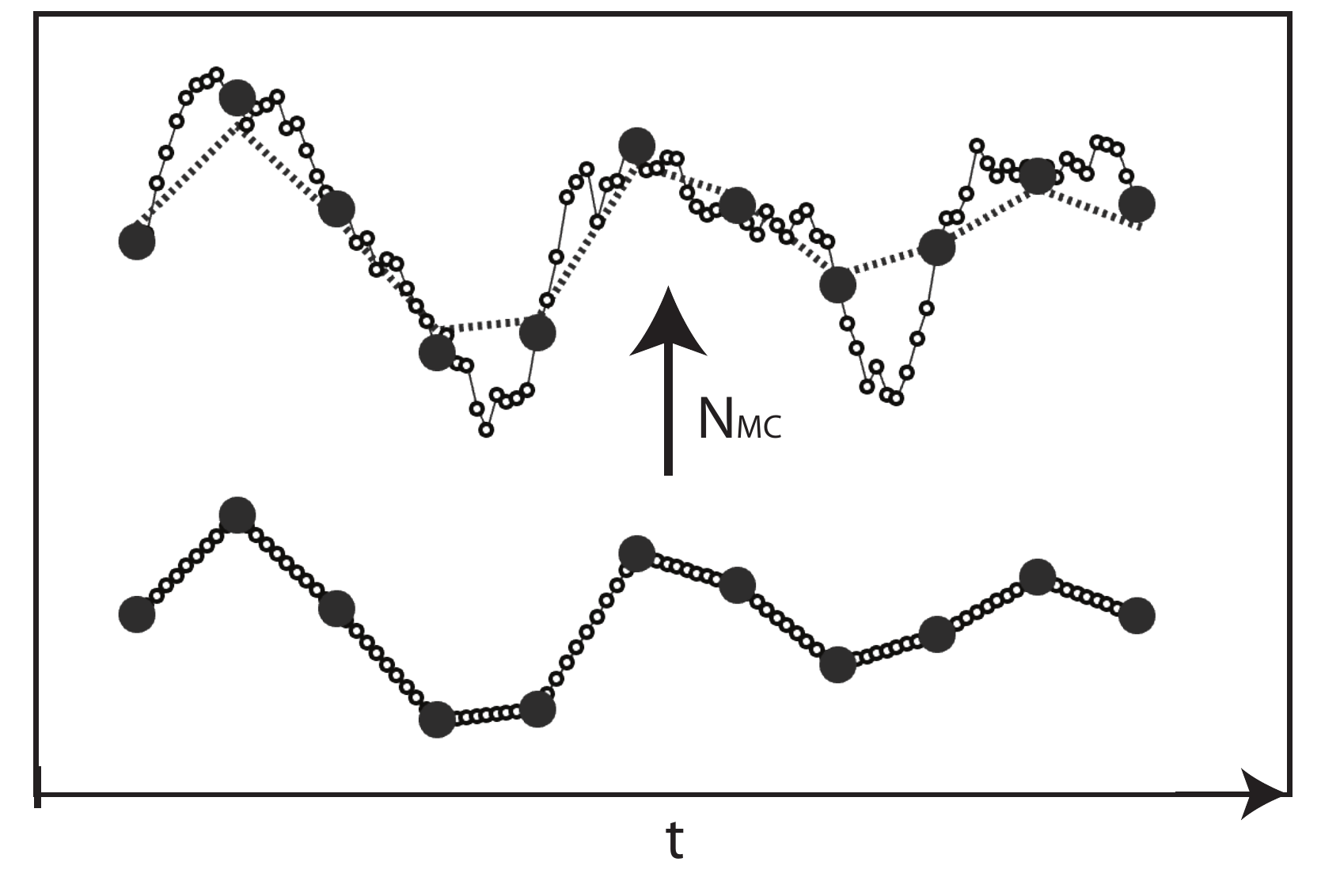}
    \caption{Simulated polymer chain dynamics, with $n+1=11$ data points (large circles) and $j-1=9$ intermediate beads (small circles). For other parameters see Sections~\ref{numerical_results} and \ref{Comp}. Bottom: initial state, where intermediate beads are on a linear interpolation between data points. Top: polymer after $N_{MC}=1000$ iterations of the propagation algorithm (dotted line: initial configuration). Clearly the new configuration is mostly determined by the dynamics of the light-mass intermediate beads, while the heavy-mass data points move to a much lesser extent.}
    \label{fig:polymer}
\end{figure}

Physically, the discretized Hamiltonian can be identified with a classical polymer chain of $N$ beads with harmonic bonds between neighboring beads in an external field \cite{chandler_1981_polymer}.
The latter consists of two parts, a field that results from the measurements and is felt by the measurement beads only (last term on the r.h.s. of Eq.~(\ref{Vdisc})), and a field that results from the dynamics of the original Eq.~(\ref{sde}) and is felt by all the beads.
The masses $m_i$ and $m_{\alpha}$ are tunable parameters of the algorithm. Since measurement beads are constrained more than intermediate beads, we will assign larger masses to the former.
Fig. (\ref{fig:polymer}) shows a typical realization of the dynamics of the polymer.
Measurement beads only move within the measurement uncertainty, whilst the intermediate beads explore much larger regions of phase space.


We have thus reduced the original Bayesian inference problem to simulating the dynamics of a linear polymer (cf. Fig.~(\ref{fig:polymer})). Each state of this fictitious molecule corresponds to a well-defined configuration in the original phase space, characterized by a set of system variables $\left\{q_i\right\}_{i=1,\dots,N}$ and a parameter vector $\bt$. It is now essential to note that potential (\ref{Vdisc}) contains terms of distinct scaling in the potentially large numbers $N$ and $n$, that refer to dynamics on distinct time-scales. In particular, for large $N$, $V(q,\bt)$ is dominated by its harmonic part, and to resolve its dynamics, brute force numerical integration of Hamilton's equations in step 2 of the HMC algorithm would require a very small discretization time-step.

\subsection{Time scale separation}\label{Sect:timescale}

Whereas an interesting {\em approximate} approach would be to employ a partial averaging of the fast Fourier modes \cite{doll_1985_fourier}, we will use an {\em exact multiple time scale integration} based on Trotter's formula \cite{tuckerman_1993}. For this, we introduce so-called {\it staging variables}, and diagonalize the harmonic part between the measurement points. To this end, we rewrite the discretized harmonic part of the action as
\begin{multline}
  \sum_{i=2}^{N}
  \frac{T}{2\Delta t}
  (q_i-q_{i-1})^2
  =
  \frac{T}{2}
  \sum_{s=1}^{n}\bigg\{
    \frac{(q_{(s-1)j+1} - q_{sj+1})^2}{j\Delta t}
    \\
    +
    \sum_{k=2}^j
    \frac{k}{(k-1)\Delta t}
    (q_{(s-1)j+k}-q^*_{(s-1)j+k})^2
  \bigg\}\,,
\end{multline}
with $  q^*_{(s-1)j+k}  =  ((k-1)q_{(s-1)j+k+1} + q_{(s-1)j+1} )/k$.
The boundary beads, corresponding to the original measurement points, are not transformed,
$  u_{sj+1} = q_{sj+1}$, $s=0,\dots,n$,
while, for the intermediate staging beads, we apply the coordinate transformations
$u_{sj+k} = q_{sj+k} - q^*_{sj+k}$, $s=0,\dots,n-1$, $k=2,\dots,j$.
Their inverse transformations are given by
\begin{align}
  q_{sj+1} &= u_{sj+1}\,,
  \\
  q_{sj+k} &= \sum_{l=k}^{j+1}\frac{k-1}{l-1}u_{sj+l}
  +\frac{j-k+1}{j}u_{sj+1}\,,
\end{align}
which can be captured by the recursive relation
\begin{equation}
  q_{sj+k} = u_{sj+k} + \frac{k-1}{k} q_{sj+k+1}+ \frac{1}{k}u_{sj+1} \,.
\end{equation}
The momenta are not transformed, which means we are using a non-canonical transformation. This alters only the dynamics of the system, not the posterior we are interested in.

We now split the Hamiltonian $\mathcal{H}_{\mbox{\tiny HMC}}$ into components according to their scaling behavior in $n$ and $N$, and write
\begin{equation}
  \mathcal{H}_{\mbox{\tiny HMC}}= \mathcal H_N+\mathcal H_n+\mathcal H_1\,,
\end{equation}
where
\begin{align}
  \mathcal H_N 
  &=
  \frac{1}{2}
  \sum_{s=1}^{n}
  \sum_{k=2}^j
  \left\{
    \frac{\Delta t}{m'}p_{(s-1)j+k}^2
    +
    \frac{Tk}{\Delta t(k-1)}
    u_{(s-1)j+k}^2
  \right\}
  \,,
  \label{H_N}
  \\
  \label{H_n}
  \mathcal H_n 
  &=
  \frac{1}{2}
  \sum_{s=1}^{n+1}
  \bigg\{
   \frac{\Delta t }{M}p_{(s-1)j+1}^2 +
    \frac{(\ln(y_s/r_{(s-1)j+1}) - {\beta u_{(s-1)j+1}})^2}{\sigma^2}
   \bigg\}
  \\
  \nonumber
  &+
  \frac{T}{2j\Delta t}
  \sum_{s=1}^{n}
    (u_{(s-1)j+1} - u_{sj+1})^2
   \,,
  \\ 
  \label{H_1}
  \mathcal H_1 &=
   \sum_{\alpha=1}^2\frac{\pi_\alpha^2}{2m_\alpha}
   +
  \frac{\Delta t}{T}
   \sum_{i=2}^{N}
   \bigg\{
    \frac{1}{2}
     \left(
        \rho_i-\frac{\beta}{\gamma}e^{-\beta q_i}
     \right)^2
    -
    \frac{\beta^2}{2\gamma}
    e^{-\beta q_i}
   -
    T q_i\dot\rho_i
   \bigg\}
   \\
  &+
  \frac{1}{\gamma}
  e^{-\beta q_N}
  +
  q_N \rho_{N}
  \nonumber
  -
  \frac{1}{\gamma}
  e^{-\beta q_1}
  -
  q_1 \rho_{2} \,.
\end{align}

Here, we have introduced two masses, $M$ and $m'$, for the boundary and staging beads, respectively.
The scaling of these Hamiltonians can be derived from basic properties of discretized SDEs, from which we conclude that $u_i\sim\sqrt{\Delta t}$. Furthermore, in agreement with the {\em equipartition law} we find that $p_i\sim 1/\sqrt{\Delta t}$.
Accordingly, we find that the harmonic part (\ref{H_N}), for the staging beads, scales linearly with $N$. The terms of Eq.~(\ref{H_n}), including both the harmonic part for the boundary beads and the measurement term, scale linearly with $n$. Finally, Eq.~(\ref{H_1}) neither scales with $n$ nor $N$.
Thanks to the staging variables, $\mathcal H_N$ and $\mathcal H_n$ have become fully decoupled.
We use Trotter's formula \cite{tuckerman_1992} in order to design a reversible molecular dynamics integrator that takes the presence of the different time scales into account.
For an appropriate partition of the Hamiltonian, three distinct regimes can be distinguished:

 \begin{enumerate}
   \item[\it (i)]
  $\mathcal H_N \sim \mathcal H_n >> \mathcal H_1$,
  \item[\it (ii)]
  $\mathcal H_N >> \mathcal H_n \sim \mathcal H_1$,
  \item[\it (iii)]
  $\mathcal H_N >> \mathcal H_n >> \mathcal H_1$.
\end{enumerate}

In the following we restrict ourselves to regime $(ii)$, where the number of measurements $n$ is assumed to be not too large and/or the measurement error $\sigma$ to be not too small (the generalization of the method to the other regimes would, however, be straightforward). In this regime we may simply separate the harmonic part of the action for the staging beads from the rest and write
\begin{equation}
  \mathcal{H}_{\mbox{\tiny HMC}}=\mathcal H_N + \mathcal H'\,.
\end{equation}
For obtaining reversible integrators, we define the Liouville operators
$ iL_N=\{\cdot\,,\,\mathcal H_N\}$, $iL'=\{\cdot\,,\,\mathcal H'\}$,
where $\{\cdot\,,\,\cdot\}$ denote the Poisson brackets that apply to functions on the phase space.
Trotter's formula \cite{trotter_1959} allows us to write the Hamiltonian propagator as
\begin{equation}\label{propagator}
  e^{i(L_N+L')\tau}
  =
  (e^{iL_N(\Delta\tau/2)}e^{iL'\Delta\tau}e^{iL_N(\Delta\tau/2)})^P
  +
  \mathcal O(\tau^3/P^{2})\,,
\end{equation}
for $\tau =P\Delta \tau$.
Here, the outer propagator $\exp[iL_N(\Delta \tau/2)]$ reflects much faster dynamics than the inner one.
However, thanks to our re-parametrization, it describes the dynamics of uncoupled harmonic oscillators, which we can readily solve.
Masses and frequencies of the oscillators are derived from (\ref{H_N}) as
\begin{equation}
  m=m'/\Delta t\,,\quad
  \omega_k=\sqrt{\frac{Nk}{(k-1)m}}\,.
\end{equation}
The outer propagator becomes
\begin{align}
  u_{(s-1)j+k}(\Delta\tau/2)
  &=
  u_{(s-1)j+k}(0)\cos(\omega_k\Delta\tau/2) +
  \frac{p_{(s-1)j+k}(0)}{m\omega_k}\sin(\omega_k\Delta\tau/2)\,,\label{analytical1}
  \\
  p_{(s-1)j+k}(\Delta\tau/2)
  &=
  p_{(s-1)j+k}(0)\cos(\omega_k\Delta\tau/2) -
  m\omega_k u_{(s-1)j+k}(0) \sin(\omega_k\Delta\tau/2)\,,\label{analytical2}
\end{align}
for $s=1,\dots,n$ and $k=2,\dots,j$.
For the inner propagator, we employ the time-reversible and volume preserving velocity Verlet algorithm \cite{swope_1982_verlet}, which leads for the boundary beads to
\begin{align}
  u_{(s-1)j+1}(\Delta\tau)
  &= u_{(s-1)j+1}(0)
  +
  \frac{\Delta\tau}{M} p_{(s-1)j+1}(0) +
  \frac{\Delta \tau^2}{2M}
  F_{(s-1)j+1}[\vc u(0),{\pmb\theta}(0)]\,,\\
  p_{(s-1)j+1}(\Delta\tau)
  &= p_{(s-1)j+1}(0)
  +
  \frac{\Delta\tau}{2}
  (
  F_{(s-1)j+1}[\vc u(0),{\pmb\theta}(0)] +
  F_{(s-1)j+1}[\vc u(\Delta\tau),{\pmb\theta}(\Delta\tau)]
  )\,,
\end{align}
with $s=1,\dots,n+1$ and where $F_i[\vc u,{\pmb\theta}]$ denotes the partial derivative of $\mathcal H'[\vc u,{\pmb\theta}]$ w.r.t. $u_i$.
Analogous equations emerge for the model parameters $\bt$ and their momenta ${\pmb\pi}$, by exchanging $u_{…}$ by $\theta_{…}$ and $p_{…}$ by $\pi_{…}$, along with the corresponding masses, respectively.
For the staging beads
only the momenta need to be updated (because the associated kinetic term is not part of $\mathcal H'$, but of $\mathcal H_N$).
Thus, with $s=1,\dots,n$ and $k=2,\dots,j$ as before,
\begin{multline}\label{propagatorStagingP}
  p_{(s-1)j+k}(\Delta\tau)
  =
  p_{(s-1)j+k}(0)
  \\
  +
  \frac{\Delta\tau}{2}
  (
  F_{(s-1)j+k}[\vc u(0),{\pmb\theta}(0)]
  +
  F_{(s-1)j+k}[\vc u(\Delta\tau),{\pmb\theta}(\Delta\tau)]
  )
  \,.
\end{multline}

The propagators (\ref{analytical1}) through (\ref{propagatorStagingP}) are applied sequentially $P$ times to calculate the system evolution over time $\tau$. The proposed configuration, $(\vc u',\bt';\vc p',{\pmb\pi}')$ is accepted with Metropolis probability $  \min
  \left(
    1,
    e^{\mathcal{H}_{\mbox{\tiny HMC}}(\vc u,\bt;\vc p,{\pmb\pi}) - \mathcal{H}_{\mbox{\tiny HMC}}(\vc u',\bt';\vc p',{\pmb\pi}')}
  \right)$.
The next iteration then starts with sampling a new momentum vector $(\vc p,{\pmb\pi})$.

The analytical solution (\ref{analytical1}) and (\ref{analytical2}) is one main boosting part of our algorithm.
To find such a solution, it is important to arrive at model equations of the form (\ref{standardform}), where the noise term neither depends on the state variables, nor on the parameters to be inferred.
In a one dimensional model this can always be achieved through re-parametrization (see, e.g., chapter 5 in \cite{risken_1989_FockerPlanck}).
In higher dimensions, this will not always be possible. But even in such cases, we will be able to boost our algorithm through assigning smaller time intervals $\Delta\tau$ to the fast dynamics and larger ones to the slow dynamics.

\section{Results}\label{numerical_results}

\begin{figure}[htb!]
    \centering
    \includegraphics[width=0.48\textwidth]{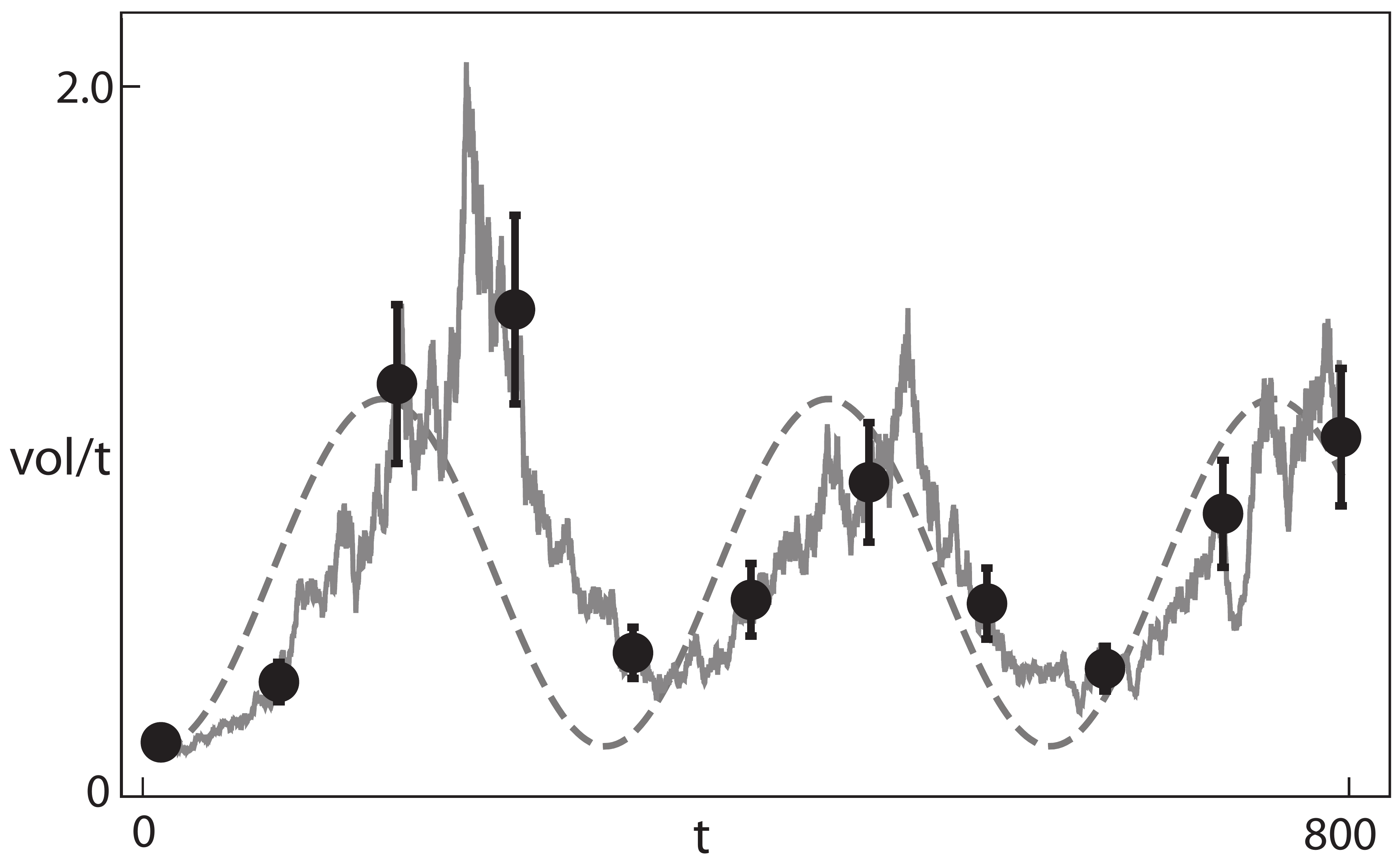}
    \caption{\label{fig:rain_data_S}
    System realization (solid line) with synthetic observations (filled circles, with error bars indicating the assumed measurement uncertainty).
    The system response closely follows the oscillations of the sinusoidal input (dashed) in a time-delayed manner.
    Parameters for this figure and all following figures: $K_{\mbox{\small true}} = 50$ and $\gamma_{\mbox{\small true}} = 0.2$.
}
\end{figure}

\begin{figure}[htb!]
    \centering
    \includegraphics[width=0.48\textwidth]{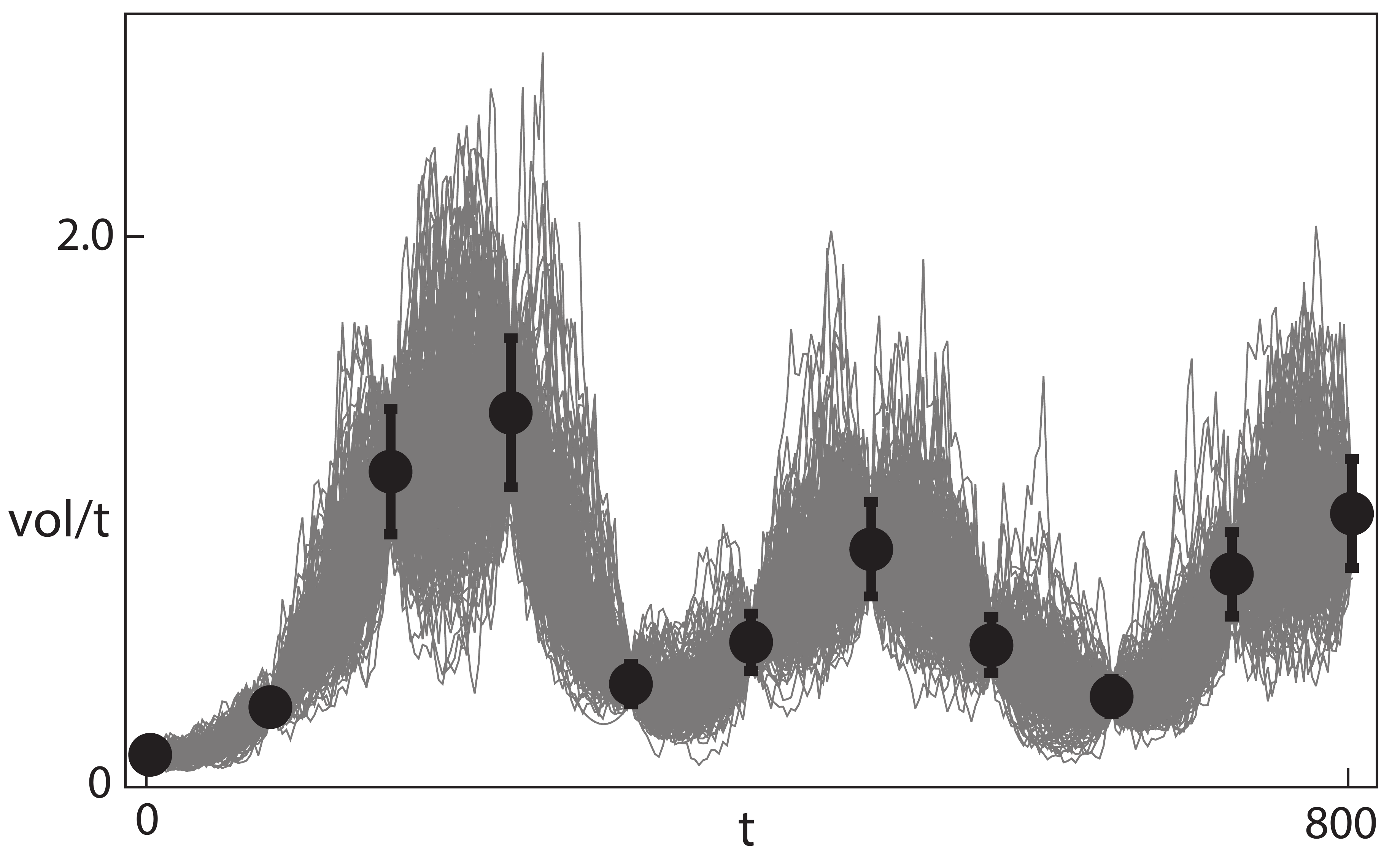}
    \caption{Simulated system realizations associated with synthetic data, based on $n+1 = 11$ measurement points and  $N=301$ discretization points.
}
    \label{fig:spaghetti}
\end{figure}

For our toy system, we have considered a simple sinusoidal input  $r(t) = \sin^2 \left( 0.01 t \right) + 0.1$. A system realization was first obtained from Eq.~(\ref{sde}) using $K_{\mbox{\small true}} = 50$ (in arbitrary units of time) and $\gamma_{\mbox{\small true}}=0.2$.
Such system realization was then used to generate a synthetic time series of observed data according to Eq.~(\ref{data}). The error $\sigma$ was set to $0.1$. The input signal, the "true" system realization and the corresponding data time series are shown in Fig.~(\ref{fig:rain_data_S}).
 A set of 200 system realizations sampled from the integrand of Eq. (\ref{posterior_pathint}), based on $n+1=11$ measurement points and $N = 301$ discretization points, is shown in Fig.~(\ref{fig:spaghetti}), together with the generated synthetic data.
These samples were generated with a HMC algorithm, where the masses were set (in arbitrary units) to $M=720$ for the measurement beads, to $m'=130$ for the intermediate discretization beads, and to $m_\alpha=150$ for both the dimensionless parameters $\beta$ and $\gamma$.
The different dynamics of the heavy measurement beads and the light discretization beads can be appreciated in Fig.~(\ref{fig:spaghetti}).

\begin{figure}[htb!]
    \centering
    \includegraphics[width=0.48\textwidth]{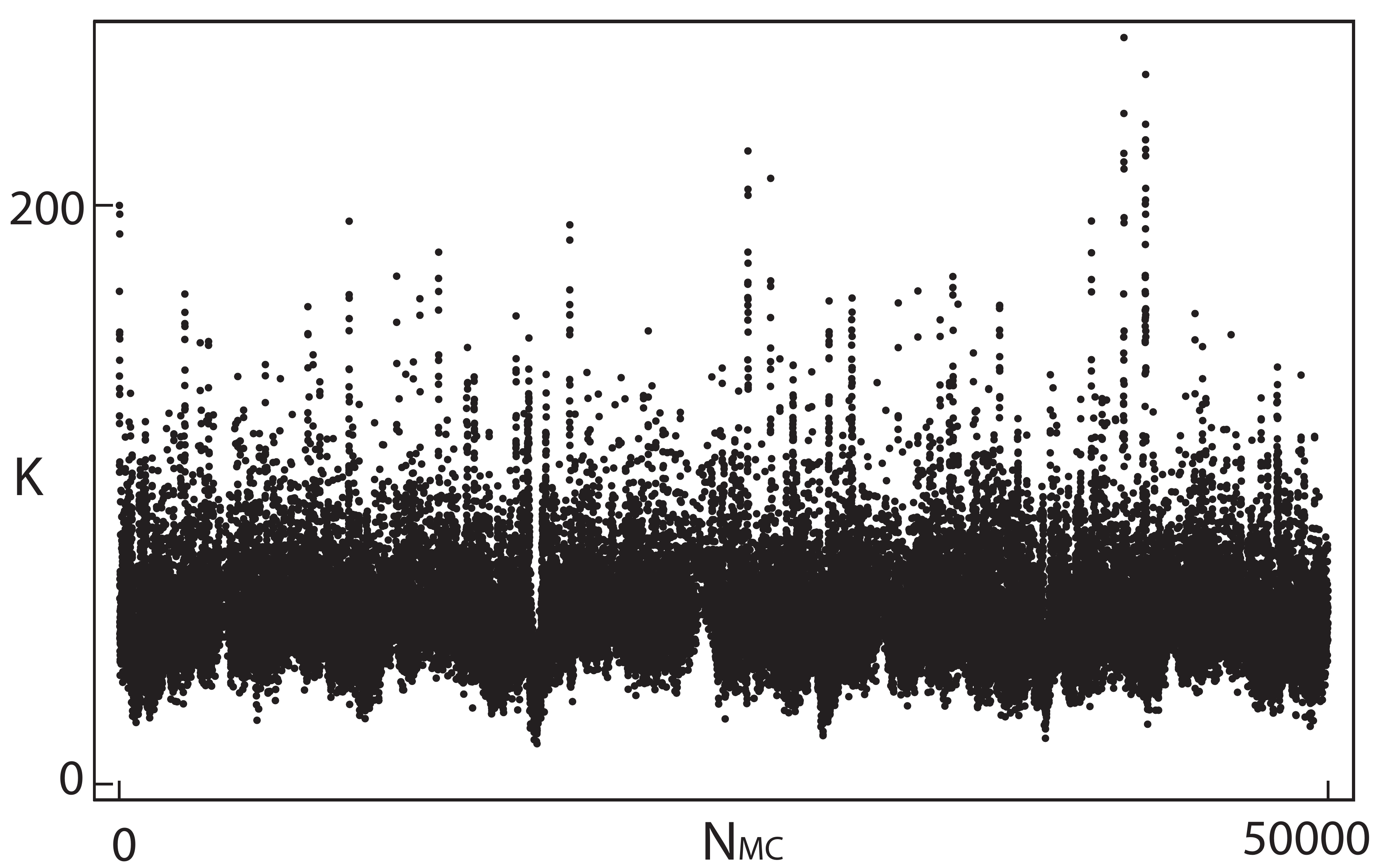}
    \caption{Markov chain evolution of the inferred parameter $K$.}
    \label{fig:chainK}
\end{figure}

\begin{figure}[htb!]
    \centering
   \includegraphics[width=0.48\textwidth]{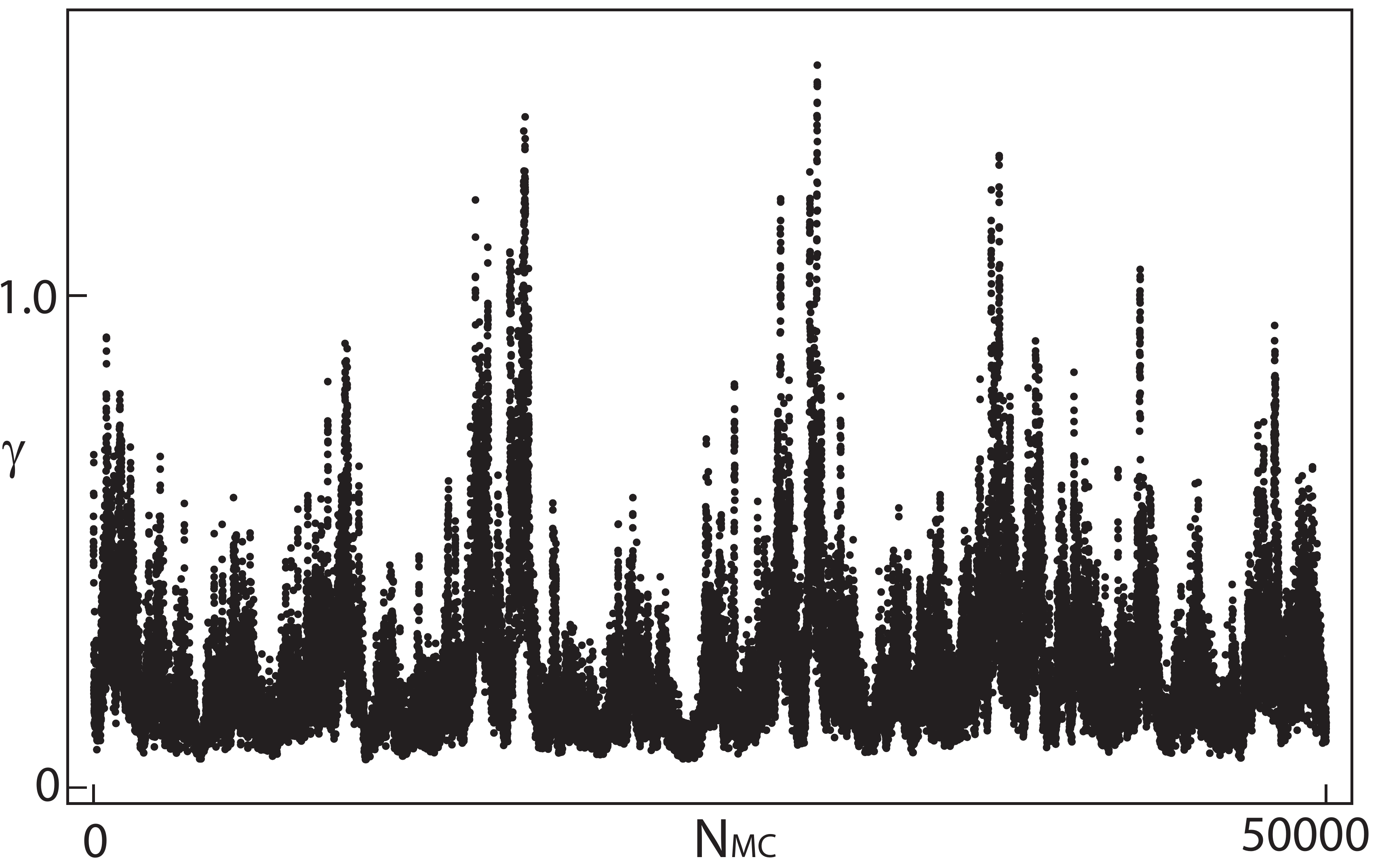}
    \caption{Markov chain evolution of the inferred parameter $\gamma$.}
    \label{fig:chainG}
\end{figure}

The Markov chains for parameters $K$ and $\gamma$, obtained after $N_{MC}=50000$ iterations of the HMC algorithm, are shown in Figs.~(\ref{fig:chainK}) and (\ref{fig:chainG}), respectively.

\begin{figure}[htb!]
    \centering
    \includegraphics[width=0.48\textwidth]{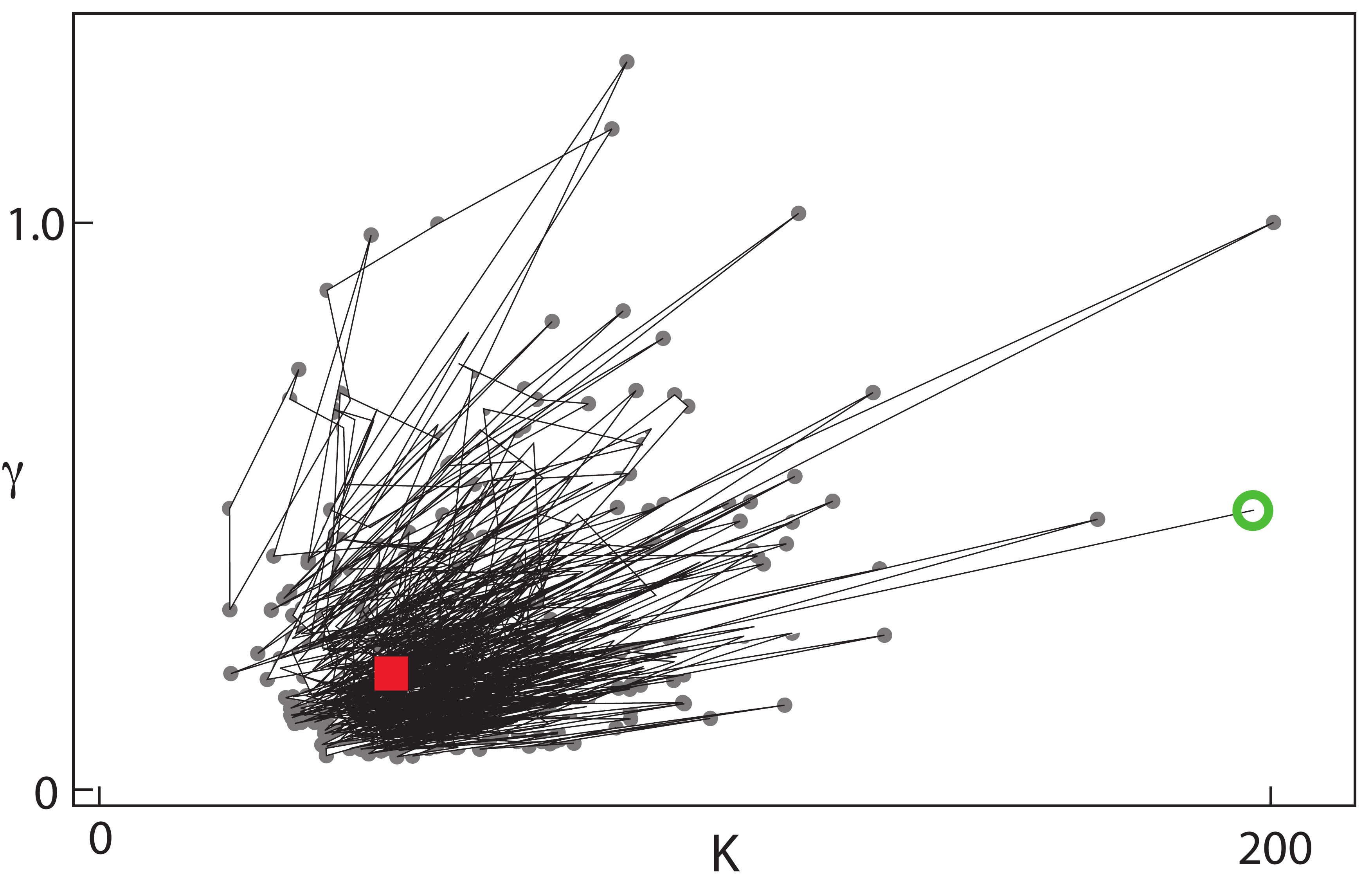}
    \caption{System dynamics in the phase space $K-\gamma$. The circle represents the initial state, while the square corresponds to the true parameter values used to generate the data.}
    \label{fig:phase_space_evol}
\end{figure}

The efficiency of the algorithm can be appreciated best by inspecting the system evolution in the phase space $K-\gamma$ (Fig.~(\ref{fig:phase_space_evol})).
The starting point of the algorithm was set to a linear interpolation of the data points, together with values for the model parameters that were deliberately chosen far off the truth (open circle in Fig. \ref{fig:phase_space_evol}).
Nevertheless, the very first step of the algorithm already takes the system to the vicinity of the true parameter values, where most of its dynamics then occurs. Few excursions lead far away from the true parameter values. These explore the heavy tails of the posterior parameter distribution.

\begin{figure}[htb!]
    \centering
    \includegraphics[width=\textwidth]{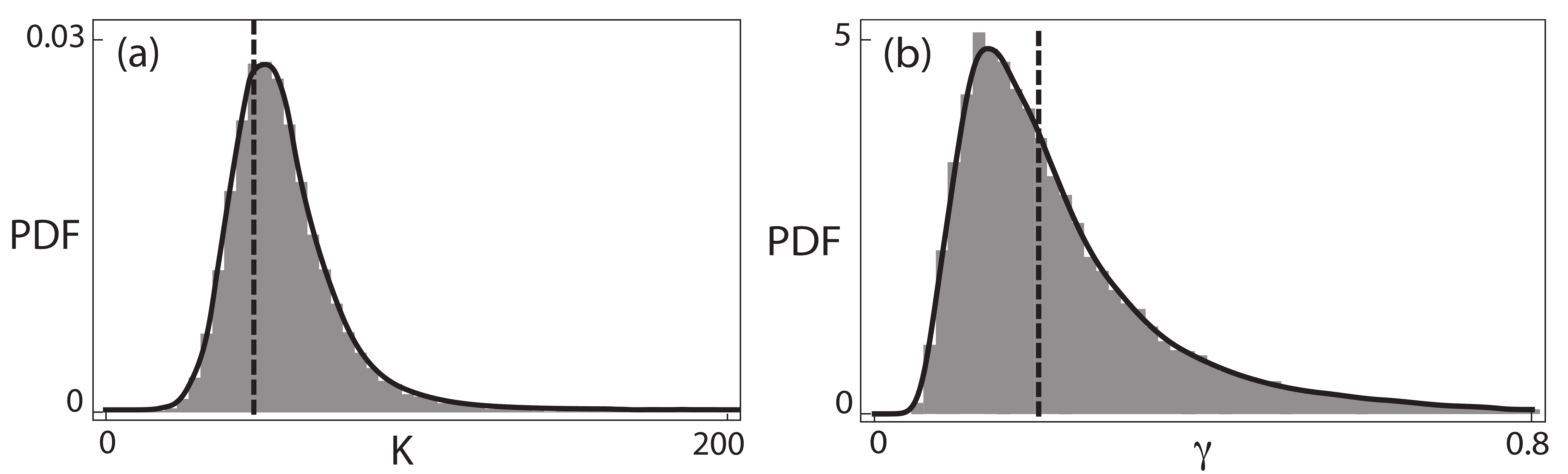}
    \caption{Probability density functions for the inferred parameters, (a) $K$ and  (b) $\gamma$. The true values used to generate the data are represented by the dashed vertical lines.}
    \label{fig:KG_distr}
\end{figure}

The Markov chains (Figs.~(\ref{fig:chainK}) and (\ref{fig:chainG})) determine the probability density functions (PDF) for $K$ and $\gamma$, respectively. The results obtained by using the built-in kernel density estimator provided by Mathematica (version 10) are exhibited in  Fig.~(\ref{fig:KG_distr}); they are fully compatible with the true, to be inferred, parameter values $K_{\mbox{\small true}}$ and $\gamma_{\mbox{\small true}}$.

\section{Implementation}\label{Comp}

The algorithm was implemented in C++ (version C++11) using the open source \emph{Adept} library (version 1.1; \cite{Hogan_2014_adept}), which provides a powerful tool for fast reverse-mode automated differentiation (AD). Our algorithm benefits greatly from the use of AD. This gives us the possibility to modify Eq.~(\ref{sde}) and therefore the action~(\ref{action}), while leaving the implementation of the algorithm unaltered. This makes our program extremely flexible and suitable for a much broader range of applications than the simple exemplary SDE model described here.

The simulations were run on both serial and parallel implementations of the algorithm on a 64-bit Linux system equipped with two 12-core 2.7~GHz processors (Intel Xeon E5-2697v2) and 64~GB of memory clocked at 1866~MHz. We used $n+1 = 11$ measurement points and $N = 101, 201, \dots, 501$ discretization points. In the Hamiltonian propagator~(\ref{propagator}), we set $\Delta\tau = 0.25$ and $P = 3$, with a constant total observation time $T = 833$ (arbitrary units of time).  The initial values of the parameters were set to $K=200$ and $\gamma = 0.5$.

For example, a complete run with $N_{MC}=50000$ iterations with $n=10$ and $N = 301$ required about 43 seconds with the serial implementation of the algorithm. In the case of our toy system the burn-in phase is extremely short and can be safely ignored. Under these conditions the algorithm can be parallelized in a straightforward manner simply by breaking up the Markov chain into several smaller independent chains. The execution time with a fixed-size problem scales in a reasonably linear way with the number of processes (strong scaling). Our example could be therefore run in only about 3 seconds using 16 processors.
An alternative strategy, suitable for long time-series, would be to parallelize the updates of the polymer beads in each step of a single MC chain.

\section{Conclusions}

We presented a novel, extremely efficient and versatile approach for data-based SDE parameter estimation.
Our algorithm obtains its strength from translating the problem of generating posterior parameter samples into the problem of simulating the dynamics of a statistical mechanics system;
the main novelty in our algorithm is the exploitation of the fact that this dynamics generically happens on very different time scales.
Furthermore, at least for 1D systems, our approach also allows for an analytical, and therefore computationally efficient, integration of the fastest part of the dynamics.

In most application cases, our choice of a fixed diagonal mass matrix, for reasonable choices of masses - heavier for the measurement beads, lighter for the discretization beads - can be expected to work well.
Nonetheless, if the curvature of the potential varies strongly, it might be beneficial to adapt the mass matrix to the local curvature as suggested in \cite{girolami_2011_HMC}. For such cases, a combination of the scale separation method proposed in this paper with the local mass matrix adaptation of \cite{girolami_2011_HMC} might be the most efficient solution. This extension, however, comes at the price of a computational overhead (second derivatives have to be calculated and implicit equations have to be solved), and we will no longer be able to solve part of the dynamics analytically.
On a more general level, given the wide field of very distinct applications, optimal parameter inference for SDE models will not be provided by one single approach, but will require a set of tools, to make the optimal choice from. We expect statistical physics to continue making strong contributions toward this aim.


The structure of our algorithm is well suited to parallelization, which is important in particular if we deal with a high number of measurements. The algorithm can easily be adapted to other inference problems, such as higher dimensional SDE, and SDE coupled to ODE. These adaptations and extensions will, however, be addressed in future works.

\section*{Funding}

This work was partly financed by the Eawag Discretionary Fund.

\bibliographystyle{plain}

\bibliography{C:/Users/albertca/SWITCHdrive/refs}

\begin{thebibliography}{10}

\bibitem{Albert_2014_ABC}
C.~Albert, H.~R. K\"unsch, and A.~Scheidegger.
\newblock {A Simulated Annealing Approach to Approximate Bayes Computations}.
\newblock {\em Stat.~Comput.}, 25(6):1217--1232, 2015.

\bibitem{alder_1959_MD}
B.~J. {Alder} and T.~E. {Wainwright}.
\newblock {Studies in Molecular Dynamics. I. General Method}.
\newblock {\em J. Chem. Phys.}, 31:459--466, 1959.

\bibitem{box_2011_bayesian}
G.~EP. Box and G.~C. Tiao.
\newblock {\em Bayesian inference in statistical analysis}, volume~40.
\newblock John Wiley \& Sons, 2011.

\bibitem{breinholt_2011_SDE}
A.~Breinholt, F.O. Thordarson, J.K. Moller, M.~Grum, P.S. Mikkelsen, and
  H.~Madsen.
\newblock {Grey-box modelling of flow in sewer systems with state-dependent
  diffusion}.
\newblock {\em {Environmetrics}}, {22}({8}):{946--961}, {2011}.

\bibitem{chandler_1981_polymer}
D.~Chandler and P.~G. Wolynes.
\newblock Exploiting the isomorphism between quantum theory and classical
  statistical mechanics of polyatomic fluids.
\newblock {\em J. Chem. Phys.}, 74(7):4078--4095, 1981.

\bibitem{chopin_2013_SMC2}
N.~Chopin, P.~E. Jacob, and O.~Papaspiliopoulos.
\newblock {SMC2: an efficient algorithm for sequential analysis of state space
  models}.
\newblock {\em {J. Roy. Stat. Soc. B}}, {75}({3}):{397--426}, {2013}.

\bibitem{doll_1985_fourier}
JD~Doll, Rob~D Coalson, and David~L Freeman.
\newblock {Fourier path-integral Monte Carlo methods: Partial averaging}.
\newblock {\em Physical review letters}, 55(1):1, 1985.

\bibitem{duane_1987}
S.~Duane, A.~D. Kennedy, B.~J. Pendleton, and D.~Roweth.
\newblock {Hybrid Monte Carlo}.
\newblock {\em Phys.~Lett.~B}, 195(2):216--222, 1987.

\bibitem{dutre_1977_SDE}
WL~Dutr{\'e} and AF~Debosscher.
\newblock {Exact Statistical Analysis of Nonlinear Dynamic Nuclear-Power
  Reactor Models by the Fokker-Planck Method—Part I: Reactor with Direct
  Power Feedback}.
\newblock {\em Nuclear Science and Engineering}, 62(3):355--363, 1977.

\bibitem{Fearnhead_2012_ABC}
P.~Fearnhead and D.~Prangle.
\newblock {Constructing summary statistics for approximate Bayesian
  computation: semi-automatic approximate Bayesian computation}.
\newblock {\em {J. Roy. Stat. Soc. B}}, {74}({3}):{419--474}, {2012}.

\bibitem{fujisaka_1986_intermittency}
H.~Fujisaka, H.~Ishii, M.~Inoue, and T.~Yamada.
\newblock Intermittency caused by chaotic modulation. ii—lyapunov exponent,
  fractal structure and power spectrum—.
\newblock {\em Progress of theoretical physics}, 76(6):1198--1209, 1986.

\bibitem{girolami_2011_HMC}
M.~Girolami and B.~Calderhead.
\newblock {Riemann manifold Langevin and Hamiltonian Monte Carlo methods}.
\newblock {\em J. Roy. Stat. Soc. B}, {73}({Part 2}):{123--214}, 2011.

\bibitem{Hogan_2014_adept}
R.~J. Hogan.
\newblock {Fast reverse-mode automatic differentiation using expression
  templates in C++}.
\newblock {\em ACM~Trans.~Math.~Softw.}, 40(4):1--16, 2014.

\bibitem{lau_2007}
A.~W.~C. Lau and T.~C. Lubensky.
\newblock State-dependent diffusion: thermodynamic consistency and its path
  integral formulation.
\newblock {\em Phys.~Rev.~E}, 76(1):011123, 2007.

\bibitem{liu_2012_filters}
X.~Liu and M.~Niranjan.
\newblock {State and parameter estimation of the heat shock response system
  using Kalman and particle filters}.
\newblock {\em Bioinformatics}, 28(11):1501--1507, 2012.

\bibitem{livina_2003_dischargeFluctuations}
V.~Livina, Y.~Ashkenazy, Z.~Kizner, V.~Strygin, A.~Bunde, and S.~Havlin.
\newblock A stochastic model of river discharge fluctuations.
\newblock {\em Physica A: Statistical Mechanics and its Applications},
  330(1):283--290, 2003.

\bibitem{marin_2012_ABC}
J.M. Marin, P.~Pudlo, C.P. Robert, and R.J. Ryder.
\newblock {Approximate Bayesian computational methods}.
\newblock {\em {Statistics and Computing}}, {22}({6, SI}):{1167--1180}, {2012}.

\bibitem{metropolis_1953}
N.~Metropolis, A.~W. Rosenbluth, M.~N. Rosenbluth, A.~H. Teller, and E.~Teller.
\newblock Equation of state calculations by fast computing machines.
\newblock {\em J.~Chem.~Phys.}, 21(6):1087--1092, 1953.

\bibitem{rahman_1964_MD}
A.~{Rahman}.
\newblock {Correlations in the Motion of Atoms in Liquid Argon}.
\newblock {\em Physical Review}, 136:405--411, 1964.

\bibitem{reichert_2009_timedepParameters}
P.~Reichert and J.~Mieleitner.
\newblock {Analyzing input and structural uncertainty of nonlinear dynamic
  models with stochastic, time-dependent parameters}.
\newblock {\em {Water Resources Res.}}, {45}, {2009}.

\bibitem{risken_1989_FockerPlanck}
H.~Risken.
\newblock {\em {The Fokker-Planck Equation; Methods of Solution and
  Applications; 2nd ed.}}
\newblock Springer, 1989.

\bibitem{rosenblatt_1958_perceptron}
F.~Rosenblatt.
\newblock The perceptron: a probabilistic model for information storage and
  organization in the brain.
\newblock {\em Psychological review}, 65(6):386, 1958.

\bibitem{schenzle_1979_multStochProc}
A.~Schenzle and H.~Brand.
\newblock Multiplicative stochastic processes in statistical physics.
\newblock {\em Phys. Rev. A}, 20(4):1628, 1979.

\bibitem{steeb_2004_problems}
W.H. Steeb, Y.~Hardy, A.~Hardy, and R.~Stoop.
\newblock {\em {Problems \& Solutions In Scientific Computing With C++ And Java
  Simulations}}.
\newblock World Scientific Publishing Co., Inc., 2004.

\bibitem{stratonovich_1968}
R.~L. Stratonovich.
\newblock {\em {Conditional Markov Processes and their Application to the
  Theory of Optimal Control}}.
\newblock Elsevier, New York, 1968.

\bibitem{swope_1982_verlet}
W.~C. Swope, H.~C. Andersen, P.~H. Berens, and K.~R. Wilson.
\newblock A computer simulation method for the calculation of equilibrium
  constants for the formation of physical clusters of molecules: Application to
  small water clusters.
\newblock {\em J. Chem. Phys.}, 76(1):637--649, 1982.

\bibitem{thyer_2009_fattails}
M.~Thyer, B.~Renard, D.~Kavetski, G.~Kuczera, S.~W. Franks, and S.~Srikanthan.
\newblock {Critical evaluation of parameter consistency and predictive
  uncertainty in hydrological modeling: A case study using Bayesian total error
  analysis}.
\newblock {\em Water~Resources~Res.}, 45(12), 2009.

\bibitem{tomassini_2009_smoothing}
L.~Tomassini, P.~Reichert, H.~R. K{\"u}nsch, C.~Buser, R.~Knutti, and M.~E.
  Borsuk.
\newblock A smoothing algorithm for estimating stochastic, continuous time
  model parameters and its application to a simple climate model.
\newblock {\em J. Roy. Stat. Soc. C}, 58(5):679--704, 2009.

\bibitem{toni_2010_ABC}
T.~Toni and M.P.H. Stumpf.
\newblock Simulation-based model selection for dynamical systems in systems and
  population biology.
\newblock {\em Bioinformatics}, 26(1):104--110, 2010.

\bibitem{toni_2009_ABC}
T.~Toni, D.~Welch, N.~Strelkowa, A.~Ipsen, and M.~P.~H. Stumpf.
\newblock {Approximate Bayesian computation scheme for parameter inference and
  model selection in dynamical systems}.
\newblock {\em {J. R. Soc. Interface}}, {6}({31}):{187--202}, {2009}.

\bibitem{trotter_1959}
H.~F. Trotter.
\newblock On the product of semi-groups of operators.
\newblock {\em Proc.~Amer.~Math.~Soc.}, 10(4):545--551, 1959.

\bibitem{tuckerman_1992}
M.~E. Tuckerman, B.~J. Berne, and G.~J. Martyna.
\newblock Reversible multiple time scale molecular dynamics.
\newblock {\em J.~Chem.~Phys.}, 97(3):1990--2001, 1992.

\bibitem{tuckerman_1993}
M.~E. Tuckerman, B.~J. Berne, G.~J. Martyna, and M.~L. Klein.
\newblock {Efficient molecular dynamics and hybrid Monte Carlo algorithms for
  path integrals}.
\newblock {\em J.~Chem.~Phys.}, 99(4):2796--2808, 1993.

\end{thebibliography}

\end{document}